%% file: acl_latex.tex
\pdfoutput=1

\documentclass[11pt]{article}

\usepackage{acl}
\usepackage{graphicx}
\usepackage{amsmath}
\usepackage{booktabs}
\usepackage{fancyvrb}
\usepackage{xcolor}         %
\usepackage{xspace}         %
\usepackage{colortbl}
\usepackage{color}

\usepackage{times}
\usepackage{latexsym}

\usepackage{todonotes}

\usepackage[T1]{fontenc}

\usepackage[utf8]{inputenc}

\usepackage{microtype}

\newcommand{\coll}{NeuCLIRBench}

\title{\coll : A Modern Evaluation Collection for Monolingual, \\
Cross-Language,
and Multilingual Information Retrieval}

\author{
    \textbf{Dawn Lawrie}$^{\hspace{.1em}\hspace{.1em}{\color{blue}\boldsymbol{\iota}}}$
    \quad
    \textbf{James Mayfield}$^{\hspace{.1em}\color{blue}\boldsymbol{\iota}}$
    \quad
    \textbf{Eugene Yang}$^{\hspace{.1em}\color{blue}\boldsymbol{\iota}}$
    \quad
    \textbf{Andrew Yates}$^{\hspace{.1em}\color{blue}\boldsymbol{\iota}}$
    \vspace{.2em}\\
    \textbf{Sean MacAvaney}$^{\hspace{.1em}\color{blue}\boldsymbol{\gamma}}$
    \quad
    \textbf{Ronak Pradeep}$^{\hspace{.1em}\color{blue}\boldsymbol{\alpha}}$
    \quad
    \textbf{Scott Miller}$^{\hspace{.1em}\color{blue}\boldsymbol{\beta}}$
    \quad
    \textbf{Paul McNamee}$^{\hspace{.1em}\color{blue}\boldsymbol{\iota}}$
    \quad
    \textbf{Luca Soldani}$^{\hspace{.1em}\color{blue}\boldsymbol{\sigma}}$
    \vspace{.5em}\\
    $^{\color{blue}\iota\hspace{.1em}}$HLTCOE, Johns Hopkins University
    \quad
    $^{\color{blue}\gamma\hspace{.1em}}$University of Glasgow
    \quad
    $^{\color{blue}\alpha\hspace{.1em}}$University of Waterloo
    \vspace{.2em}\\
    $^{\color{blue}\gamma\hspace{.1em}}$Information Sciences Institute, University of Southern California
    \quad
    $^{\color{blue}\gamma\hspace{.1em}}$Allen Institute for AI
    \vspace{.5em}\\
    \texttt{\{lawrie, mayfield, eugene.yang, andrew.yates\}@jhu.edu}
    \\}

\begin{document}

\maketitle

\begin{abstract}

To measure advances in retrieval,
test collections with relevance judgments that can faithfully distinguish systems are required.
This paper presents \coll,
an evaluation collection for cross-language and multilingual retrieval.
The collection consists of documents written natively in Chinese, Persian, and Russian,
as well as those same documents machine translated into English.
The collection supports several retrieval scenarios including:
monolingual retrieval in English, Chinese, Persian, or Russian;
cross-language retrieval with English as the query language
and one of the other three languages as the document language;
and multilingual retrieval,
again with English as the query language and relevant documents in all three languages.
\coll\ combines the TREC NeuCLIR track topics of 2022, 2023, and 2024.
The 250,128 judgments across approximately 150 queries for the monolingual and cross-language tasks
and 100 queries for multilingual retrieval provide strong statistical discriminatory power
to distinguish retrieval approaches.
A fusion baseline of strong neural retrieval systems
is included with the collection
so that developers of reranking algorithms are no longer reliant on BM25 as their first-stage retriever.
\coll{} is publicly available.\footnote{\url{https://huggingface.co/datasets/neuclir/bench}}

\end{abstract}

\input{1_intro}

\input{2_related_work}

\input{3_dataset_creation}

\input{4_results}

\input{5_conc}

\bibliography{custom}

\input{6_appendix}

\end{document}

%% file: 1_intro.tex
\section{Introduction}

\begin{figure*}[ht]
    \centering
    \includegraphics[width=\linewidth]{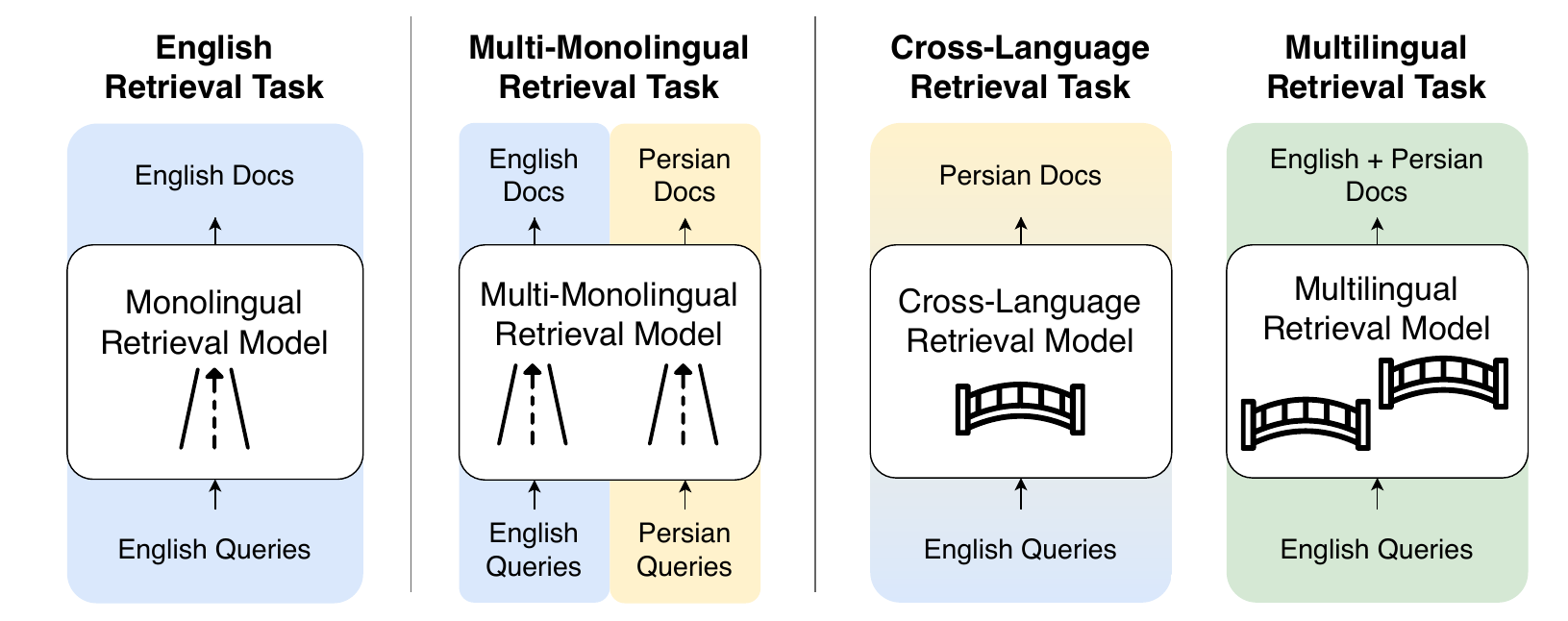}
    \caption{NeuCLIR Tasks}
    \label{fig:tasks}
\end{figure*}

This paper presents \coll, a retrieval test collection over newswire documents
that enables evaluation of future first-stage retrieval systems and rerankers
for several retrieval tasks over four diverse languages,
as shown in Figure~\ref{fig:tasks}.
A retrieval test collection consists of \textit{documents}, \textit{queries},
and \textit{relevance judgments} indicating which documents are relevant to each query.
The 10 million documents in \coll{} are drawn from natively written Chinese, Persian, and Russian articles
that appeared in CommonCrawl News, 
providing a challenging scenario for first-stage retrieval systems to index given its size.
To support the evaluation of English monolingual retrieval,
all documents have been translated into English using machine translation.
Cross-Language Information Retrieval (CLIR) evaluation in this paper uses English queries
and documents in the other three languages.
The collection includes manual translations of the queries into Chinese, Persian and Russian;
these queries support multi-monolingual experiments,
as well as cross-language retrieval experiments over other language pairs 
(\textit{e.g.,} Chinese queries with Russian documents).
Finally, the collection supports evaluation of Multilingual Retrieval approaches
by combining the documents from all three languages and using queries expressed in English.

\coll{} is derived from the work of the TREC Neural Cross-Language Information Retrieval (NeuCLIR) track,
which ran from 2022 to 2024.
\coll{} combines the queries from the three years of that track.
With over 100 queries per task, there is strong statistical discriminatory power
when trying to distinguish retrieval approaches.
\coll{} provides deep judgments for each query so that
nearly all relevant documents for a query were judged by a human assessor.
It includes a fusion baseline for reranking that combines three modern retrieval approaches:
a single-vector dual-encoder, a multi-vector dual-encoder, and a learned-sparse approach.
Until now, most rerankers have been evaluated using a BM25 first-stage retriever,
exposing powerful rerankers only to documents that have lexical matches with the query.
This fusion baseline allows rerankers to be tested on their ability to rank highly 
relevant documents that lack lexical matches with the query.

The remainder of the paper compares \coll{} to popular test collections used for these tasks (Section~\ref{sec:rw});
summarizes the main aspects of test collection creation (Section~\ref{sec:build});
and provides results for monolingual English retrieval,
multi-monolingual retrieval,
cross-language retrieval,
and multilingual retrieval (Section~\ref{sec:results}).

%% file: 2_related_work.tex
\section{Related Work}
\label{sec:rw}

There has long been an interest in information retrieval systems capable of processing multiple languages,
 and therefore a longstanding need for corresponding test collections.
 As an early example, TREC-4 built a Spanish test collection for adhoc retrieval
allowing the same system evaluated on both English and Spanish~\cite{DBLP:conf/trec/Harman95}.
 Later, TREC-6 introduced a cross-language retrieval task~\cite{DBLP:conf/trec/SchaubleS97}
 that included three distinct document corpora
 (English, German, and French)
 with a unified set of topics translated into each language.
 This setting enabled benchmarking of both monolingual and cross-language retrieval.
 Shortly thereafter, benchmarking efforts in non-English monolingual and cross-language retrieval
 were taken up by other evaluation campaigns,
 most notably CLEF~\cite{DBLP:conf/clef/HarmanBHKPSS00},
 NTCIR~\cite{DBLP:conf/ntcir/KandoKNEKH99},
 and FIRE~\cite{DBLP:conf/fire/MajumderPBM10}.
Information retrieval evaluation campaigns eventually focused primarily on monolingual settings
and other information access tasks.
The cross-language test collections from these campaigns are generally of high-quality,
given their use of native bilingual speakers for annotation.
However, they still have several problems that motivate our new benchmark:
they are comparatively small by modern standards
(e.g., TREC-4's Spanish collection contains only 120K documents,
compared to the millions typically used in modern evaluations);
the labels are not necessarily complete enough to measure today's retrieval systems;
and there is a risk of data contamination
since Large Language Models are very likely trained on them~\cite{DBLP:journals/corr/abs-2411-02284}.

A recent trend in information retrieval evaluation is to benchmark monolingual retrieval
in many languages at once,
without aligning the queries or the documents~\cite{mrtidy, miracl, bge-m3}.
We call such datasets \textit{multi-monolingual},
since they are a collection of many distinct monolingual datasets
rather than a true multilingual or cross-language dataset.
One example is MIRACL~\cite{miracl},
which uses Wikipedia articles in 18 languages
and generates queries by asking annotators to create questions that cannot be answered in the first 100 words of an article.
Although these benchmarks are useful for measuring how well retrieval systems perform in many individual languages,
they cannot be used to evaluate \textit{cross-language} retrieval systems
(i.e., with a query written in one language and documents in another).

A common technique relies on machine translation to fill this gap,
usually by translating the document corpus of an existing collection from one language into another~\cite{DBLP:journals/corr/abs-2108-13897}.
This approach is limited for several reasons.
First, machine-translated content is not representative of naturally-occurring content,
both in terms of language characteristics (``\textit{translationese}'')
and in the topicality of the content itself~\cite{DBLP:conf/emnlp/GrahamHK20}.
Furthermore, translated datasets are not necessarily fit as evaluation data
since no human has verified that the translated content is still relevant to the queries.
Nevertheless, machine translation plays an important role in training cross-language retrieval systems~\cite{DBLP:conf/ecir/YangLMOM24}.

Yet another trend is to use ``found pairs'' of bilingual text to construct cross-language retrieval benchmarks~\cite{DBLP:conf/emnlp/SunD20,DBLP:conf/acl/SchamoniHSR14}.
For instance, CLIRMatrix~\cite{DBLP:conf/emnlp/SunD20} uses cross-language WikiData links between articles
to construct queries from article titles in one language and article content from another language.
Although this process yields natural text,
using document titles as the source of query text causes the dataset to model only basic entity search behavior.

In summary, existing human-annotated cross-language retrieval test collections
do not meet modern information retrieval evaluation standards of scale and task complexity,
while other techniques
(machine translation and found text)
face issues with the realism of the text content or the task complexity.
Collectively, these limitations motivate us to introduce \coll,
a modern cross-language and multilingual retrieval benchmark.

%% file: 3_dataset_creation.tex
\section{Building the Datasets}
\label{sec:build}

The \coll{} test collection has three components:
documents, topics, and relevance judgments.
We describe each of these components in turn. 

\subsection{Documents}
\label{sub:documents}

\begin{table*}[tb]
\centering
\caption{Document Collection Statistics for NeuCLIR-1.
Token counts from SpaCy~\cite{spacy}.}\label{tab:coll_stats}
\begin{tabular}{cccccc}
\toprule
 & \multicolumn{1}{c}{Document}  & \multicolumn{1}{c}{Avg. Chars}  & \multicolumn{1}{c}{Median Chars}  & \multicolumn{1}{c}{Avg. Tokens}  & \multicolumn{1}{c}{Median Tokens}  \\ 
Language &  \multicolumn{1}{c}{count} & \multicolumn{1}{c}{per Document} & \multicolumn{1}{c}{per Document} &  \multicolumn{1}{c}{per Document} & \multicolumn{1}{c}{per Document} \\ 
\midrule
Chinese & 3,179,209 & 743 & 613 & 427 & 356 \\
Persian & 2,232,016 & 2032 & 1427 & 429 & 300 \\
Russian & 4,627,543 & 1757 & 1198 & 301 & 204 \\ 
\bottomrule
\end{tabular}
\end{table*}

The document collection, NeuCLIR-1, consists of documents in three languages:
Chinese, Persian, and Russian,
drawn from CommonCrawl News.\footnote{\url{https://commoncrawl.org/2016/10/news-dataset-available/}}
The documents were obtained by the CommonCrawl service between August 1, 2016 and July 31, 2021;
most of the documents were published within this five-year window.
The documents are released as text files without HTML markup.
Text was extracted from each source web page using the Python utility \textit{Newspaper}.\footnote{\url{https://github.com/codelucas/newspaper}}

The collection is distributed as JSONL,
a list of JSON objects, one per line.
Each line represents a document.
Each document JSON structure consists of the following fields:
\begin{description}
\itemsep0em 
    \item [id:] UUID assigned by CommonCrawl 
    \item [cc\_file:] CommonCrawl source file containing the document %
    \item[timestamp:] time of publication (or, if unknown, CommonCrawl download time)
    \item[title:] article headline or title
    \item[text:] article body
    \item[url:] address of the source web page
\end{description}

To ascertain the language of each document,
its title and text were independently run through two automatic language identification tools: \textit{cld3}\footnote{\url{https://pypi.org/project/pycld3/}},
a neural language identification tool;
and \textit{VALID}~\cite{mcnamee-2016-language}, a compression-based model trained on Wikipedia text.
Documents for which the tools agreed on the language,
or where one of the tools agreed with the language recorded in the web page metadata,
were included in the collection under the language of agreement;
all others were removed. 
This is an imperfect process and some documents comprising text in other languages are included in the collection.
The extent of such language pollution is unknown;
however, annotators sometimes encountered out-of-language documents in the pools of documents to be assessed.
Such documents were always considered to be not relevant.
All documents with more than 24,000 characters (approximately ten pages of text) were also removed,
as very long documents create assessment challenges.
Very short documents were also removed, specifically:
Chinese documents containing 75 or fewer characters;
Persian documents containing 100 or fewer characters;
and Russian documents containing 200 or fewer characters.
We observed that such documents are often not genuine news articles,
frequently consisting of isolated headlines or commercial advertisements.
All remaining documents were deduplicated so that only one instance of each document remained in the collection.

Each collection was limited in size to at most five million documents.
After removing duplicates, the Russian collection was significantly above this threshold.
Therefore, we used scikit-learn's implementation of random sampling without replacement\footnote{\url{https://scikit-learn.org/stable/modules/generated/sklearn.utils.random.sample_without_replacement.html}}
to downsample the collection.
Final collection statistics appear in Table~\ref{tab:coll_stats}.

Since a timestamp is included for each document,
the collection can be ordered to support streaming experiments such as in~\citet{plaid-shirttt}

\subsection{Queries}
\label{sec:topics}

Queries were developed in batches during the three years that the NeuCLIR Track ran at TREC.
All years used the same general development process,
which consisted of an ideation phase, 
a search phase where documents were examined and judged based on relevance,
and a revision phase.
There were minor variations in the process over the years
based on the language skills of the query creators
and the tools available during the search phase. 

\coll{} query creators were instructed to develop traditional TREC-style information needs.
These are broader than the queries found in most question-answering evaluation collections,
which can be answered with a phrase or sentence.
TREC information needs,
usually called \textit{topics,}
consist of a short title,
a sentence-length description,
and a paragraph-length narrative.
During ideation, query creators write a description indicating what is being searched for,
and a narrative that discusses what is and is not considered to be relevant to the topic.
During the search phase the query creator uses a search engine to identify a set of documents to judge.
The creator counts the number of relevant documents in the first search results.
Queries with too many relevant documents are revised or discarded.
During the revision phase, the description and narrative may be altered to reflect nuances in document judgment.
Finally, the query creator chooses a title that summarizes the topic in three to five words.

All topics coming out of this phase were reviewed by the Track Coordinators;
topics that were overly ambiguous, too similar to another topic,
or judged by the organizers to be otherwise inappropriate were eliminated.
The remaining topics were distributed to track participants.
Track submissions were then used during relevance judgment,
as discussed in Section~\ref{sec:rel-judge}.

As alluded to above,
slight changes were made to query creation over the three years of the program.
Queries numbered 0 to 199 were each created by a single bilingual query creator.
During the search phase, the creator used monolingual search in the language of the collection. 
The creator counted the number of relevant documents in the top 30 documents ranked using BM25.
Any query that had between 1 and 20 relevant documents was added to the set of queries to be reviewed by the coordinators.

Queries numbered 200 to 247 were created by a pair of bilingual query creators,
each of whom had language skills in a different document language. 
In a virtual meeting, the creators brainstormed a topic together.
Their instructions were that good topics must ``revolve around events, people, and places,
and be significant enough to have coverage in more than one language.''
After exploring the collection with monolingual searches in the two languages,
they initiated a single monolingual search in each language.
Each creator counted the number of relevant documents in the top 25 documents in their language,
ranked using BM25.
Creators were instructed to revise the topic if the count was zero or greater than 20 in either of the languages.

Queries numbered 248 to 299 were developed by the Track Coordinators,
most of whom lacked language skills in any of the collection's non-English languages.
Query creators worked independently to craft topics
using the same general three-step process used to create the earlier topics.
However, the tools available to them were different.
Since these creators entered queries in English rather than in the language of the collection,
they were given access to two CLIR search engines:
a BM25-based engine that indexed machine translated documents;
and a dense retrieval system implemented with ColBERT-X~\cite{colbert-x} that indexed the original documents.
The creators' document display included both the original document and the machine translation.
The browser could also be used to invoke Google Translate on the original document
should the user desire to do so.   
They could also use HiCAL~\cite{hical} to recommend documents to be judged.
Creators made relevance assessments using the same relevance categories:\footnote{Defined in Section~\ref{sec:rel-judge}}
``very valuable,'' ``somewhat valuable,'' ``not that valuable,'' and ``not relevant.''
They were asked to judge at least thirty documents
and find between one and twenty documents in the very or somewhat valuable categories.
Document judgments were recorded and added to the judgment pools for final relevance assessments. 

Queries 300 to 399 followed a similar approach to that of Queries 200 to 247. 
A pair of bilingual query creators with skills in different document languages worked together on query ideation.
However, rather than judging documents ranked by BM25,
their monolingual query returned documents ranked by ColBERT-X~\cite{colbert-x}.
The switch was made because the BM25 counts did not reliably identify queries with too many relevant documents.

\subsection{Relevance Judgments}
\label{sec:rel-judge}

Each year, TREC NeuCLIR track participants submitted the results of their system runs to NIST.
Each query creator with bilingual language skills served as a relevance assessor.
The assessors read a subset of the retrieved documents,
called the \textit{judgment pool},
in the original language of the document.
Thus, each query was assessed by three different assessors when it was judged in all three languages.
Pool composition varied slightly over the query sets,
but always consisted of at least 20 documents per run.
This ensured complete judgments for all runs when using the nDCG@20 metric,
the primary metric used to evaluate systems.
In the following we describe how the judgment pools were assembled and how relevance was determined.

\subsubsection{Creating Judgment Pools}

Pools were created from the submitting systems' top-ranked documents. 
For Queries 0 to 199,
the number of documents that a run contributed to a pool
was based on whether the submitting team marked the run as a baseline run
(see \citet{lawrie2022overview} for a list of these runs).
Baseline runs contributed their top 25 documents,
while non-baseline runs contributed their top 50 documents.
For Queries 200 to 299,
the top 50 documents for runs that teams prioritized as their top three runs were added to the pools
(see \citet{lawrie2023overview} for a list of these runs).
For other runs, a depth of twenty was used.
For Queries 300 to 399,
the top 100 documents for runs that teams prioritized as their top nine runs were included in the pools
(see \citet{lawrie2024overview} for a list of these runs). 
For other runs, a depth of 50 was used.

\subsubsection{Relevance Judgment Process}

Assessors used a four-point scale to judge the relevance of each document.
Assessors were instructed to provide their assessment
as if they were gathering information to write a report on the topic.
Relevance was assessed on the most valuable information found in the document;
the grades and their associated graded relevance scores were:
\begin{description}
\itemsep0em 
\item[Very Valuable:] (3 points) information that would appear in the lead paragraph of a report on the topic
\item[Somewhat Valuable:] (1 point) information that would appear somewhere in a report on the topic
\item[Not that Valuable:] (0 points) information that does not add new information beyond the topic description,
or information that would appear only in a report footnote
\item[Not Relevant:] (0 points) a document without any information about the topic
\end{description}

\subsubsection{Query Analysis}

During the assessment periods, 167 Chinese queries, 174 Persian queries,
and 171 Russian queries were judged.
Within each language some topics had fewer than three relevant documents,
while other topics had a concerningly large percentage of relevant documents in the pools.
Having topics with fewer than three relevant documents
can have undesirable effects on the ability to statistically distinguish systems;
they have been removed from \coll. %
There are 16 Chinese topics, 27 Persian topics, and 16 Russian topics
with fewer than three relevant documents. 
Thus each language has at least 147 topics in \coll.

The requirements for the MLIR task in \coll{} are slightly different.
Instead of requiring at least three relevant documents in each language,
the requirement is applied across all three languages;
thus, it is possible for a query to have no relevant documents in a particular language
as long as the query was assessed in that language
and the other two languages combine to have a sufficient number of documents.
Only one query was dropped for having too few relevant documents;
this resulted in 105 topics for the monolingual English and the multilingual retrieval tasks.

Identifying and eliminating topics likely to contain many unidentified relevant documents is important
for future use of the collection.
One approach simply calculates the percentage of relevant documents in the pool
and sets a cutoff (such as 20\% prevalence)
above which we cannot be confident that the relevant set is sufficiently identified.
Using this cutoff would flag 10 topics in Chinese, 9 topics in Persian, and 14 topics in Russian.
Another way of investigating this issue is by examining the occurrence of unjudged documents
among those used to calculate the score
(in this case those that are in the top 20).
The only documents that could possibly affect the score of a system
are unjudged documents that are relevant but are treated as non-relevant.
As long as the percentage of judged documents in the top results remains high for future systems,
those systems will be scored fairly.
If that percentage drops significantly,
it is not possible to determine whether systems are being scored fairly without additional judgments.
We encourage users of the collection to report judged@20 averages when reporting results.

\subsection{Resources}
\coll{} relies on the NeuCLIR-1 document set, which comprises documents written in Chinese, Persian, and Russian.
The collection also includes updated document translations into English.
Documents were re-translated using a vanilla Transformer model
that was trained in-house with the \textit{Sockeye} version 2 toolkit~\cite{sockeye2}
using recently released bitext obtained from publicly available corpora.\footnote{\url{https://opus.nlpl.eu}}
The number of sentences used in training is given in Table \ref{tbl:bitext},
along with BLEU scores on the FLORES-200 and NTREX-128 benchmarks for each language \cite{flores200,ntrex128}.

Canonicalized query files are in tab-separated format
with the query ID followed by the query. 
Queries in \coll{} are the concatenation of the title and description. %
Each query file is written in a single language.
All queries were first expressed in English,
then manually translated into the document languages
to produce queries that naturally express the meaning of the original query in those languages.
An additional query file is also released in JSONL format
containing the narratives written to describe document relevance,
as well as translations of the titles, descriptions, and narratives using GoogleTranslate
(translated in the summer of the year it appeared at TREC).
These translations support CLIR experiments where the underlying retrieval approach
expects queries in the same language as the documents; thus, they are a means for crossing the language barrier.

The relevance judgments are released in four qrels\footnote{\url{https://trec.nist.gov/data/qrels_eng/}} files,
one per language plus one covering all three languages.
The non-English ones can be used for CLIR and multi-monolingual experiments;
the MLIR file is used for monolingual English and multilingual retrieval experiments. 

\begin{table}
\centering
\caption{MT training data used and BLEU scores on FLORES-200 and NTREX-128 benchmarks.}
\label{tbl:bitext}
\begin{tabular}{lrrr}
\toprule
Language & $\#$ Sents & FLORES & NTREX \\
\midrule
Chinese & 127M & 32.7 & 29.3 \\
Persian & 26M &  39.4 & 30.5 \\
Russian & 194M & 36.2 & 29.4 \\
\bottomrule
\end{tabular}
\vspace{-0.5em}
\end{table}

%% file: 4_results.tex
\input{tables/mono}

\section{Evaluation on NeuCLIRBench}
\label{sec:results}

In this section, we provide a comprehensive set of results from commonly-reported retrieval models in prior works.
The full set of model descriptions can be found in Appendix~\ref{sec:models-desc}. 
These models include dense and sparse neural bi-encoders and rerankers. 
While this is not a complete list of all publicly available multi-monolingual, cross-language, and multilingual retrieval models,
it includes representatives of each type of model
as well as the state-of-the-art effectiveness of each as reported in the literature.

\subsection{Multi-Monolingual Retrieval}

Multi-monolingual retrieval results over the four languages are summarized in Table~\ref{tbl:mono}. 
The top section of the table reports the results for first-stage retrieval,
while the bottom section reports results of reranking the strong Fusion baseline.
The fusion run is a combination of PLAID-X~\cite{yang2024distillation},
MILCO~\cite{nguyen2025milco},
and Qwen3 8B Embed~\cite{qwen3embedding}
using reciprocal rank fusion~\cite{cormack2009reciprocal}.
These three models are representatives of multi-vector dense,
learned-sparse,
and single-vector dense retrieval models that are also multilingual.
We include other models in the table for reference. 
The effectiveness of the fusion is the highest reported for the first stage retrievers,
indicating that the reranking task is challenging as it starts with a very strong ranking.

The reranking results are in the bottom section of the Table,
which includes a number of popular and state-of-the-art pointwise and listwise rerankers. 
Several rerankers, such as Qwen3 0.6B Rerank~\cite{qwen3embedding},
Jina Reranker~\cite{jinav3},
and RankZephyr~\cite{pradeep2023rankzephyr},
fail to improve the initial fusion ranking in both English and the multi-monolingual average
(the ``avg'' column).  
Despite strong effectiveness reported on other benchmarks,
which mostly rerank BM25 results,
these models fail to improve a very strong initial ranking;
this indicates that more development is needed for reranking models to distinguish relevant documents
from closely related non-relevant documents. 

\input{tables/clir_mlir}

A majority of these runs have Judged@20 values higher than 0.9;
even the zero-shot RankQwen-32B model is above 0.95.
In the worst case, we observe these numbers dipping slightly below 0.8. 
These high judgment rates at the top of the rank indicate strong reusability of NeuCLIRBench on all languages,
especially since the majority of the reported models were developed later
and were not part of the pooling when judgments were created.

\subsection{Cross-Language and Multilingual Retrieval}

We report results in Table~\ref{tbl:clir-multi}
on the same set of retrieval models for the cross-language and multilingual retrieval tasks. 
While these results exhibit similar trends,
the detailed model ordering is slightly different,
providing interesting insights
For example, Qwen3 8B Reranker is more effective than its 4B variant in the monolingual task across all four languages,
although they are similar in the cross-language tasks.
The 4B variant is even higher than the 8B one in the multilingual retrieval task,
indicating that the effectiveness gain from using a larger model observed in prior work
might not hold when multiple languages are involved in the same task.

Multilingual retrieval is an even more challenging task for the models,
as the differences between models are smaller.
While Rank1~\cite{weller2025rank} is noticeably more effective than any other pointwise reranker in the cross-language task,
it is barely different from Qwen3 4B Reranker in the multilingual task.
The gap between Rank1 (the best pointwise reranker) and Rank-K (the best listwise reranker)
is also smaller in the multilingual task than in the cross-language task,
also indicating that ranking documents in different languages is more challenging.

Since most multilingual models are trained monolingually on many languages,
it is unclear how well these models would perform when the task involves multiple languages,
such as cross-language and multilingual retrieval. 
In fact, based on our reported evaluation results,
multilingual retrieval is noticeably harder for multilingual models. 
We challenge the community to explore more challenging retrieval and multilingual scenarios
to further drive the development of multilingual modeling and retrieval.

%% file: tables/mono.tex
\begin{table*}[t]
\centering
\caption{Multi-Monolingual Retrieval Results. 
Since SPLADEv3 is an English-only model we report its results on only the English retrieval task. 
Models in each group are sorted by the average nDCG@20.}
\label{tbl:mono}

\small
\setlength{\tabcolsep}{5.5pt}
\begin{tabular}{l|cccc|c|cccc}
\toprule
{}                    & \multicolumn{5}{c|}{nDCG@20} & \multicolumn{4}{c}{Judged@20} \\
{}                    & Chinese & Russian & Persian & English & Avg & Chinese & Russian & Persian & English \\
\midrule

BM25           & 0.391 & 0.408 & 0.391 & 0.349 & 0.385 & 1.00 & 1.00 & 1.00 & 0.97 \\
\midrule
\multicolumn{10}{l}{Bi-Encoders} \\
\midrule

RepLlama              & 0.326 & 0.335 & 0.110 & 0.350 & 0.280 & 0.73 & 0.79 & 0.31 & 0.91 \\
e5 Large              & 0.323 & 0.309 & 0.348 & 0.293 & 0.318 & 0.69 & 0.69 & 0.68 & 0.80 \\
BGE-M3 Sparse         & 0.277 & 0.289 & 0.386 & 0.326 & 0.319 & 0.71 & 0.65 & 0.77 & 0.85 \\
Qwen3 0.6B Embed      & 0.481 & 0.431 & 0.434 & 0.408 & 0.439 & 0.89 & 0.88 & 0.80 & 0.95 \\
PLAID-X               & 0.461 & 0.450 & 0.490 & 0.388 & 0.447 & 0.88 & 0.90 & 0.89 & 0.77 \\
SPLADEv3              & -- & -- & -- & 0.420 & -- & -- & -- & -- & 0.97 \\
MILCO                 & 0.420 & 0.461 & 0.494 & 0.423 & 0.449 & 0.78 & 0.87 & 0.82 & 0.90 \\
Arctic-Embed Large v2 & 0.470 & 0.454 & 0.493 & 0.414 & 0.458 & 0.89 & 0.89 & 0.88 & 0.96 \\
JinaV3                & 0.477 & 0.469 & 0.495 & 0.398 & 0.460 & 0.89 & 0.88 & 0.87 & 0.94 \\
Qwen3 4B Embed        & 0.541 & 0.519 & 0.550 & 0.431 & 0.510 & 0.94 & 0.94 & 0.93 & 0.96 \\
Qwen3 8B Embed        & 0.543 & 0.526 & 0.565 & 0.429 & 0.516 & 0.95 & 0.95 & 0.94 & 0.96 \\
\textit{Fusion}       & 0.577 & 0.564 & 0.609 & 0.483 & 0.558 & 0.98 & 0.98 & 0.97 & 0.99 \\
\midrule
\multicolumn{10}{l}{Pointwise Rerankers on \textit{Fusion}} \\
\midrule
Qwen3 0.6B Rerank     & 0.545 & 0.513 & 0.507 & 0.423 & 0.497 & 0.98 & 0.98 & 0.96 & 0.99 \\
Jina Reranker         & 0.565 & 0.506 & 0.504 & 0.468 & 0.511 & 0.97 & 0.97 & 0.96 & 0.98 \\
SEARCHER Reranker     & 0.534 & 0.519 & 0.512 & 0.482 & 0.512 & 0.92 & 0.95 & 0.92 & 0.99 \\
Mono-mT5XXL           & 0.572 & 0.553 & 0.592 & 0.468 & 0.546 & 0.98 & 0.99 & 0.98 & 0.99 \\
Qwen3 4B Rerank       & 0.569 & 0.580 & 0.611 & 0.494 & 0.563 & 0.98 & 0.98 & 0.97 & 0.99 \\
Qwen3 8B Rerank       & 0.595 & 0.597 & 0.620 & 0.511 & 0.581 & 0.98 & 0.98 & 0.97 & 0.99 \\
Rank1                 & 0.647 & 0.604 & 0.629 & 0.503 & 0.596 & 0.95 & 0.95 & 0.95 & 0.98 \\
\midrule
\multicolumn{10}{l}{Listwise Rerankers on \textit{Fusion}} \\
\midrule
RankZephyr 7B         & 0.522 & 0.432 & 0.552 & 0.435 & 0.485 & 0.97 & 0.95 & 0.97 & 0.99 \\
FIRST Qwen3 8B        & 0.545 & 0.598 & 0.615 & 0.518 & 0.569 & 0.97 & 0.98 & 0.97 & 0.99 \\
RankQwen-32B          & 0.640 & 0.613 & 0.657 & 0.517 & 0.607 & 0.98 & 0.98 & 0.98 & 0.99 \\
Rank-K (QwQ)          & 0.653 & 0.642 & 0.659 & 0.539 & 0.623 & 0.97 & 0.98 & 0.98 & 0.99 \\

\bottomrule
\end{tabular}

\vspace{-0.5em}
\end{table*}

%% file: tables/clir_mlir.tex
\begin{table*}[t]
\centering
\caption{Cross-Language and Multilingual Retrieval Results.
While it is possible to concatenate the translation of the queries from all three languages to support BM25 with query translation
(BM25 w/ QT)
in the multilingual setting,
it would not be a fair comparison as the length of the query is tripled.
For consistency, we still report BGE-M3 Sparse here,
but it is not designed for scenarios where the queries and documents do not share the same vocabulary. 
Models in each group are ordered by the nDCG@20 on the multilingual retrieval task.}
\label{tbl:clir-multi}

\small
\setlength{\tabcolsep}{4.5pt}
\begin{tabular}{l|ccc|c|ccc||c|c}
\toprule
{}                    & \multicolumn{7}{c||}{\textbf{Cross-Language Retrieval}} & \multicolumn{2}{c}{\textbf{Multilingual Retrieval}} \\
{}                    & \multicolumn{4}{c|}{nDCG@20} & \multicolumn{3}{c||}{Judged@20} 
                      & nDCG@20 & Judged@20 \\
{}                    & Chinese & Russian & Persian & Avg & Chinese & Russian & Persian & &  \\
\midrule

BM25 w/ QT            & 0.365 & 0.384 & 0.380 & 0.376 & 1.00 & 1.00 & 1.00 & -- & -- \\
BM25 w/ DT            & 0.439 & 0.400 & 0.447 & 0.429 & 0.96 & 0.97 & 0.97 & 0.349 & 0.97 \\
\midrule
\multicolumn{9}{l}{Bi-Encoders} \\
\midrule
BGE-M3 Sparse         & 0.089 & 0.051 & 0.044 & 0.061 & 0.24 & 0.20 & 0.16 & 0.054 & 0.30 \\
e5 Large              & 0.269 & 0.296 & 0.302 & 0.289 & 0.59 & 0.68 & 0.64 & 0.205 & 0.70 \\
RepLlama              & 0.353 & 0.391 & 0.133 & 0.292 & 0.78 & 0.83 & 0.44 & 0.255 & 0.86 \\
Qwen3 0.6B Embed      & 0.468 & 0.433 & 0.366 & 0.422 & 0.86 & 0.86 & 0.74 & 0.309 & 0.92 \\
Arctic-Embed Large v2 & 0.435 & 0.445 & 0.446 & 0.442 & 0.83 & 0.87 & 0.82 & 0.352 & 0.92 \\
JinaV3                & 0.434 & 0.468 & 0.464 & 0.455 & 0.82 & 0.86 & 0.81 & 0.355 & 0.93 \\
MILCO                 & 0.431 & 0.476 & 0.494 & 0.467 & 0.81 & 0.88 & 0.81 & 0.395 & 0.93 \\
PLAID-X               & 0.495 & 0.463 & 0.529 & 0.495 & 0.93 & 0.95 & 0.93 & 0.396 & 0.98 \\
Qwen3 4B Embed        & 0.546 & 0.526 & 0.544 & 0.538 & 0.93 & 0.92 & 0.91 & 0.415 & 0.97 \\
Qwen3 8B Embed        & 0.551 & 0.545 & 0.568 & 0.555 & 0.94 & 0.94 & 0.92 & 0.423 & 0.97 \\
\textit{Fusion}       & 0.573 & 0.581 & 0.617 & 0.590 & 0.98 & 0.99 & 0.98 & 0.468 & 0.99 \\
\midrule
\multicolumn{9}{l}{Pointwise Rerankers on \textit{Fusion}} \\
\midrule
Qwen3 0.6B Rerank     & 0.534 & 0.481 & 0.452 & 0.489 & 0.97 & 0.97 & 0.94 & 0.368 & 0.99 \\
Jina Reranker         & 0.521 & 0.473 & 0.415 & 0.469 & 0.96 & 0.96 & 0.92 & 0.373 & 0.98 \\
Mono-mT5XXL           & 0.550 & 0.536 & 0.568 & 0.551 & 0.98 & 0.98 & 0.97 & 0.445 & 0.99 \\
SEARCHER Reranker     & 0.578 & 0.544 & 0.559 & 0.560 & 0.95 & 0.96 & 0.94 & 0.468 & 0.99 \\
Qwen3 8B Rerank       & 0.599 & 0.582 & 0.587 & 0.590 & 0.98 & 0.98 & 0.97 & 0.475 & 0.99 \\
Qwen3 4B Rerank       & 0.582 & 0.583 & 0.605 & 0.590 & 0.98 & 0.98 & 0.97 & 0.487 & 0.99 \\
Rank1                 & 0.649 & 0.618 & 0.619 & 0.628 & 0.95 & 0.96 & 0.94 & 0.488 & 0.99 \\
\midrule
\multicolumn{9}{l}{Listwise Rerankers on \textit{Fusion}} \\
\midrule
RankZephyr 7B         & 0.552 & 0.441 & 0.568 & 0.520 & 0.98 & 0.96 & 0.97 & 0.424 & 0.99 \\
FIRST Qwen3 8B        & 0.544 & 0.591 & 0.602 & 0.579 & 0.98 & 0.98 & 0.97 & 0.449 & 0.99 \\
RankQwen-32B          & 0.634 & 0.625 & 0.661 & 0.640 & 0.98 & 0.99 & 0.98 & 0.480 & 0.99 \\
Rank-K (QwQ)          & 0.655 & 0.636 & 0.671 & 0.654 & 0.98 & 0.98 & 0.98 & 0.506 & 0.99 \\

\bottomrule
\end{tabular}

\vspace{-0.5em}
\end{table*}

%% file: 5_conc.tex
\section{Conclusion}

NeuCLIRBench synthesizes the test collection creation work done as part of TREC NeuCLIR 2022-2024.
It provides a new test set for evaluating cross-language, multi-monolingual, and multilingual retrieval systems
with a large, recent document collection covering three diverse languages,
many queries with relevant documents in each of those languages,
and relevance judgments we believe to cover almost all relevant documents.
It includes high quality machine translations of all documents into English,
and manual translations of the English queries into the three document languages.
It also includes a new baseline run that fuses three strong first-stage retrievals,
providing a strong starting point for rerankers.
We believe these features make \coll{} durable enough
to facilitate a wide variety of retrieval experiments for many years to come.

%% file: 6_appendix.tex
\appendix
\section*{Appendix}
\section{Model Descriptions}\label{sec:models-desc}

\subsection{Sparse Retrieval}

Sparse retrieval methods perform lexical matches between the query and the document.
Based on the lexical matches, a score is produced for each retrieved document.
We conduct experiments with BM25 using the Patapsco framework~\cite{costello22-patapsco}. 
Patapsco is a framework built on top of Pyserini~\cite{pyserini} specifically designed for experiments across many languages.
Given the reliance on lexical matches, the input query and documents must be in the same language. 
Some experimental settings such as tokenenization and token normalization are language-specific,
while others are language agnostic.
The language agnostic settings include the BM25 hyperparameter settings
where $k1=0.9$ and $b=0.4$ following standard conventions. 
The RM3 query expansion technique~\cite{rm3} 
added ten words based on the top ten documents,
weighting terms from the original query and the expansion terms equally.

For the English retrieval task,
the spaCy tokenizer and stemmer~\cite{spacy} are used to process the queries and documents,
which have stopwords removed.
For the multi-monolingual retrieval task,
all characters were normalized.
spaCy was used in all languages to perform tokenization.
For Chinese, no stemming was applied.
For Russian, the spaCy stemmer was used,
while for Persian, the Parsivar~\cite{parsivar} stemmer was used. 
For the cross-language task,
machine translation was used to cross the language barrier.
BM25-DT indicates that the documents are translated into English and then the English setting is used.
BM25-QT uses machine translation to translate the query into the language of the documents,
then the mononlingual setting for that language is followed.
In the multilingual task, only the document translation approach is applicable.
Query translation produces several versions of the query
and scores from the different languages are not comparable;
thus these scores cannot be used to rank documents across languages.

\subsubsection{Multi-Dense Vector Retrieval}

We use PLAID-X~\cite{DBLP:conf/ecir/YangLMOM24,yang2024distillation},
a cross-language and multilingual variant of the ColBERT model.
PLAID-X is fine-tuned from an XLM-RoBERTa Large model
with cross-language distillation from an mT5 reranker (described below)
using MS-MARCO v1 training queries and passages.
Each document is separated into passages using a sliding window of size 180 tokens with a stride of 90.
At inference time, document scores are aggregated from the passage scores using MaxP~\cite{maxp}.

\subsubsection{Learned Sparse Retrieval}

Learned Sparse Retrieval (LSR) methods represent queries and documents as sparse vectors
where each dimension is associated with a term in a vocabulary.
Some LSR methods produce vectors containing only terms from the input text,
while others use a Masked Language Modeling head to also perform term expansion~\cite{nguyen2023unified}.
To efficiently use these representations for first-stage retrieval,
they can be stored in an inverted index.
We conduct experiments with several LSR methods using Anserini~\cite{yang2017anserini},
which is built on Lucene.

\begin{itemize}
    \item{BGE-M3-Sparse}~\cite{bge-m3} is an LSR method trained as part of the multi-granularity M3-Embedding model. It is based on an XLM-RoBERTa backbone with 0.6B parameters. This method does not perform expansion, so it can only be used for monolingual and multi-monolingual retrieval.
    \item{MILCO}~\cite{nguyen2025milco} is a multilingual LSR method that performs both query expansion and document expansion. MILCO performs multilingual retrieval by producing sparse vectors that are tied to an English vocabulary regardless of the input language, so the model implicitly translates non-English inputs. It is based on an XLM-RoBERTa backbone with 0.6B parameters.
    \item{SPLADE-v3}~\cite{lassance2024splade} is a monolingual LSR method that performs both query expansion and document expansion. It is trained on only English MS MARCO data~\cite{bajaj2018ms}, so it can only be used for monolingual retrieval in English.
\end{itemize}

\subsubsection{Dense Retrieval}
Dense Retrieval (DR) methods represent queries and documents as a single dense vector
in a latent space~\cite{lin2022pretrained}.
To efficiently use these representations for first-stage retrieval,
they can be stored in a vector index that supports approximate nearest neighbor search,
such as FAISS~\cite{douze2025faiss}.
We use fast scan product quantization with FAISS where the number of codes is set to half the embedding dimensionality
and codes are 4 bits each.

\begin{itemize}
    \item Arctic Embed v2~\cite{yu2025arcticembed} is a multilingual model based on a XLM-RoBERTa backbone with 0.6B parameters. We use the \texttt{snowflake-arctic-embed-l-v2.0} checkpoint, which is the largest available Arctic Embed model.
    \item E5~\cite{wang2024multilingual} is a multilingual model based on an XLM-RoBERTa backbone with 0.6B parameters. We use the \texttt{multilingual-e5-large-instruct} checkpoint.
    \item Jina v3~\cite{jinav3} is a multilingual model based on an XLM-RoBERTa backbone with 0.6B parameters.
    \item Qwen3-Embedding~\cite{qwen3embedding} is a family of embedding models trained with synthetic data on top of Qwen3 backbones~\cite{yang2025qwen3}. We use all three model sizes: 0.6B, 4B, and 8B.
    \item RepLlama~\cite{rankllama} is a dense retrieval model based on LLaMA-2-7B. We use the \texttt{repllama-v1-7b-lora-passage} checkpoint trained on MS MARCO passages.
\end{itemize}

\subsubsection{First-Stage Retrieval Fusion}

We use reciprocal rank fusion (RRF)~\cite{cormack2009reciprocal} to combine the rankings
produced by three strong first-stage retrieval models from the three different families described above:
PLAID-X, MILCO, and Qwen-Embedding 8B.
Following common practice and \citet{cormack2009reciprocal}, we set $k=60$ and do not use per-method weights.

\subsubsection{Rerankers}

We rerank the top-$100$ results from the first-stage retrieval fusion
using a range of pointwise and listwise reranking models.
Pointwise models take a query-document pair as input and output a relevance score.
Listwise models take a query and set of documents as input.
Some listwise models output one relevance score for each document,
whereas others output an ordering of their input documents without producing relevance scores.
For models that output an ordering,
we use the partial sorting algorithm from RankGPT~\cite{sun2023chatgpt} to produce a ranking of all 100 results,
where we rerank 20 documents at a time with a sliding window stride of 10 from the bottom of the top 100 candidate documents.

\begin{itemize}
    \item FIRSTQwen-8B~\cite{zijian2024first} is a listwise reranking model built on top of Qwen3 8B~\cite{yang2025qwen3} and trained with the FIRST objective~\cite{reddy2024first}.
    \item jina-reranker-v3~\cite{wang2025jinarerankerv3lateinteractiondocument} is a reranking model built on top of Qwen3 0.6B that can perform pointwise or listwise reranking. In both configurations, the model outputs one relevance score for each document. We use it in a pointwise configuration.
    \item mT5~\cite{bonifacio2022mmarco} is a pointwise reranking model built on top of a multilingual T5 backbone. We use the \texttt{mt5-13b-mmarco-100k} checkpoint with 13B parameters.
    \item Qwen3-Reranker~\cite{qwen3embedding} is a family of pointwise reranking models built on top of Qwen3. We use all three model sizes: 0.6B, 4B, and 8B.
    \item Rank1~\cite{weller2025rank} is a pointwise reranking model fine-tuned to perform reasoning before reranking. Rank1 is built on top of the Qwen2.5 base model~\cite{qwen2.5}.
    \item Rank-K~\cite{yang2025rank} is a listwise reranking model that performs reasoning before reranking using the QwQ 32B reasoning model~\cite{qwq32b}. 
    \item RankQwen-32B is a zero-shot listwise reranking system that uses Qwen3-32B\footnote{\url{https://huggingface.co/Qwen/Qwen3-32B}} without any training to rerank the top-100 documents. %
    This system \emph{does not} use any reasoning.
    \item RankZephyr-7B~\cite{pradeep2023rankzephyr} is a listwise reranking model built on top of Mistral 7B~\cite{jiang2023mistral}. 
    \item SEARCHER~\cite{searcher2024} is a pointwise reranking model built on top of \texttt{Mistral-Nemo-Base-2407}.\footnote{\url{https://huggingface.co/mistralai/Mistral-Nemo-Base-2407}} It was trained on Persian, Russian, and Chinese translations of MS MARCO passages~\cite{bajaj2018ms}, keeping the queries in English. Documents in the training batch were a mixture of all three languages.
\end{itemize}